\documentclass[12pt]{article}
\usepackage{epsfig}
\begin{document}
\thispagestyle{empty}
\begin{flushright}
MCTP-00-011\\
\end{flushright}
\vspace{0.5cm}
\begin{center} 
{\Large \bf 
On the Resummation of Large QCD Logarithms in $H\to\gamma\gamma$}\\[3mm]
\vspace{1.7cm}
{\sc \bf R. Akhoury, H. Wang and  O. Yakovlev}\\[1cm]
\begin{center} \em 
Michigan Center for Theoretical Physics\\
Randall Laboratory of Physics\\ 
University of Michigan, Ann Arbor, MI 48109-1120, USA 
\end{center}\end{center}
\vspace{3cm}
\begin{abstract}
We study the strong corrections to the Higgs coupling to two photons. 
This coupling is the dominant mechanism for Higgs production in 
photon-photon collisions. In addition, the two photon decay mode 
of the Higgs is an important and  
relatively background free channel of relevance at the LHC and the Tevatron. 
We develop a method for the resummation of large QCD corrections
in the form of Sudakov-like logarithms of the type 
$\alpha_s^p\ln^{2p}({m^2 \over m^2_H})$ 
and $\alpha_s^p\ln^{2p-1}({m^2 \over m^2_H})$ (where $m$ is the light
quark mass)  which can contribute to this process in certain models 
(for example, the MSSM for large $tan \beta$) up to
next-to-leading-logarithmic (NLL) accuracy.  The NLL correction 
is  moderate, the substantial   
part of which comes from terms not related to running coupling effects.  
\end{abstract}

\vspace*{\fill}
\noindent { e-mail: akhoury, haibinw, yakovlev@umich.edu}

\newpage
\section{Introduction}
The origin of electroweak symmetry breaking is one of the central issues
in particle physics. Within the Standard Model (SM), the solution to this 
problem is associated with
the Higgs mechanism, which predicts a fundamental 
neutral scalar Higgs particle. The Higgs boson is the only SM elementary particle 
which has not been detected thus far. The precision electroweak measurements suggest 
the existence of a light Higgs boson in the mass region \cite{LEPHIGGS}
\begin{eqnarray}
113.5 < m_H < 200\quad\mbox{GeV}.
\end{eqnarray} 
In this letter we discuss the Higgs-$\gamma \gamma$ vertex,
which is extremely important for Higgs physics, specially for
Higgs boson masses in the range mentioned above. 
First, it is the main
mechanism for Higgs production in photon-photon collisions 
\cite{Ginz1,Ginz2,Telnov}; second, it is
a relatively background free decay mode for the Higgs  
at the LHC and the Tevatron and finally, the coupling is sensitive to new physics, 
and can be considered to be a counter of the number of new heavy particles.
Because of these last two observations the Higgs-$\gamma\gamma$ coupling should be 
very well calibrated. Thus, for example, the radiative corrections 
to this process should be under control. 

The coupling of the Higgs boson to two photons is absent at 
tree level in the Standard Model. The first non-zero contribution 
arises from  fermions and W boson loops. Because the Yukawa coupling 
of the quark is proportional to the quark mass, 
the contributions of the light quarks as well as charm and 
bottom quarks are well suppressed in comparison to the top quark loop
contribution.  The radiative corrections are well studied in this case:  
there is some literature devoted to the QCD and electroweak  
radiative corrections \cite{MeYa1,MeYa2,ZeSp1,KoYa,MeStHqq}. 
The only source of QCD corrections at the two loop order is the gluon
corrections to the top quark loop.  
The QCD corrections are small for Higgs masses,  
$m_H < 2m_t$, as shown by the explicit
calculations in  \cite{ZeSp1,MeYa1}. 
For heavy Higgs masses, $m_H > 2 m_t$, the corrections are large
(about 40\%), although this mass range appears to be ruled out by the 
electroweak data. 

It was observed in \cite{MeYa1} that in the limit of large 
$\frac{m_H}{m}$, where $m$ is the quark mass, the 
Higgs-$\gamma\gamma$ form factor gives the QCD double  logarithmic asymptotic, 
\begin{eqnarray}
F=F^{1-loop}(1-\frac{\rho}{12}),
\end{eqnarray} 
with $\rho
=\frac{C_F\alpha_s(\mu^2)}{2\pi}\ln^2\Big(\frac{m_H^2}{m^2}\Big)$. 
For such a factor to be 
phenomenologically relevant one must consider the contribution
to the form factor from light quark loops which as mentioned above are 
suppressed in the Standard Model. 
In fact, the $b$ quark contribution is  only about 
$4\%$ of that of the top quark at 
lowest nontrivial order and decreases with radiative
corrections. However, the contributions of the bottom quark are 
enhanced by $tan \beta $ in the super-symmetric extensions of the
Standard Model, and 
therefore the limit $m^2_H/m^2 \gg 1$, which we consider in this  paper
may be phenomenologically relevant quite apart from its inherent
theoretical interest. 
Current estimates place allowed values of $ tan \beta $ in a 
wide range \cite{tanbeta}. For large values of $tan \beta \sim 30 $, which are allowed, 
the bottom quark contributions to the Higgs-$\gamma\gamma$ form factor 
can become comparable to that of the top quark in such models. 
 
Super-symmetric extensions of the Standard Model have been extensively discussed in the
literature partly because of the possibility of resolving the hierarchy problem.
The Minimal Super-symmetric Standard Model (MSSM) stabilizes the mass of the light Higgs 
bosons in the presence of high energy GUT scales. In the MSSM, spontaneous symmetry 
breaking is induced by two complex Higgs doublets leading to five Higgs particles, 
including light and heavy scalar [CP-even] particles h and H; a
pseudo-scalar [CP-odd] particle A, and a pair of charged Higgs 
particles \cite{MSSM}. The lightest of these is predicted to 
be below $m_Z$, although radiative corrections 
 increase this limit up to 130 GeV due to large top
quark contributions \cite{HiggsRC,tanbeta}. 
 We have such models in mind, wherein as discussed above
the $b$ quark and $t$ quark contributions to the Higgs-$\gamma\gamma$ form factor 
may become comparable. Then in order to understand the radiative corrections
to the form factor it is  important to better understand these
large logarithmic corrections mentioned above and to find a method for resumming them.  

The nature of the leading double logarithmic (DL) terms of the form
$\alpha_s^p \ln^{2p}({m_H \over m})$ has been clarified
in \cite{FKM, KoYa}.  It was shown that these logarithms are
related to Sudakov form factor which enters into the one loop triangle
diagram. In this letter we discuss how the resummation of these large
logarithms to the next-to-leading-logarithmic (NLL) accuracy may be carried out. 
As will be seen, the next to leading contribution is moderately small compared 
to the leading one, implying that the radiative corrections 
to the Higgs-$\gamma\gamma$ form factor are 
under control for the case (like for example the MSSM for large $tan \beta $ ) 
when the light quark contributions cannot be neglected.

This paper is organized as follows: In section 2  
we discuss the resummation procedure of the leading double logarithms, and
in section 3 we suggest a method for  resumming the next to leading logarithmic
contributions as well. Section 4 contains the numerical results and 
discussions. In section 5 we give our conclusions.
 
\section{The Method}

As mentioned earlier the Born coupling H-$\gamma \gamma$ 
is zero in the SM as well as in the MSSM. 
At the one loop level, we find that the coupling is mediated by any
charged heavy particle loop. In the case of the SM,
the only contributions come from the quark loops and W-boson loops.
The amplitude of the Higgs decay into two photons through quark loop can
be presented in the form 
\begin{eqnarray}
M(H\to\gamma\gamma)=(e^*_1)_\mu (e^*_2)_\nu d^{\mu\nu}(G_F\sqrt{2})^{1/2}
\frac{\alpha}{4\pi}N_c\sum_{q}e_q^2 F_Q(t),
\end{eqnarray}
here $k_1$ and $k_2$ are the momenta of the photons, 
$e_1, e_2$ are the corresponding polarization vectors,
and $s=(k_1+k_2)^2=m_H^2$, $k_1^2=k_2^2=0$. 
The structure $d^{\mu\nu}=(k_1\cdot k_2)g^{\mu\nu}-k^\mu_2k_1^\nu$   
can be constructed using QED gauge invariance and Lorenz invariance 
of the amplitude.
The partial width is
\begin{eqnarray}
\Gamma (H\to\gamma\gamma )=\frac{G_F\alpha m_H^3}{128\sqrt{2}\pi^3}
\Big( \sum_{Q} N_C e_Q^2g_QF_Q +g_WF_W \Big)^2,
\end{eqnarray} 
with \cite{Ellis,Shifman}
\begin{eqnarray}
F_Q= 2t^{-1}((1-t^{-1})f(t)+1), \qquad F_W=-t^{-1}\Big( 3+2t+3(2-t^{-1})f(t)\Big)
\end{eqnarray}
and,
\begin{eqnarray}
f(t)= -\frac{1}{4}\Big(\ln
\frac{\sqrt{t}+\sqrt{t-1}}{\sqrt{t}-\sqrt{t-1}} - 
i\pi \Big)^2\quad\mbox{at}\quad t=\Big( \frac{m_H}{2m}\Big)^2 > 1.
\end{eqnarray}

In this letter we focus on the quark loop contributions only. These contributions 
will be common to all models considered. We wish to identify the leading and the
next to leading logarithmic corrections and to present a procedure 
for resumming them.

The one loop DL correction arises when the quark line opposite the
 $Hq\bar q$ vertex is soft and this can be easily evaluated using the 
Sudakov parametrization \cite{Sudakov,Gorshkov}; namely we decompose the 
soft quark momenta in terms of those along the hard photon momenta $k_1,
 k_2$  and transverse to it 
\begin{eqnarray}
l=\alpha k_1 +\beta k_2 + l_{\bot},\qquad s=(k_1+k_2)^2=m_H^2.
\end{eqnarray}
The DL contribution comes only from the region  
\begin{eqnarray}\label{kinem1}
m^2, |l_{\bot}|^2 \ll s|\alpha|,s|\beta| \ll s.
\end{eqnarray}
The loop integration in terms of the new variables reads 
\begin{eqnarray}
\int\limits_{-\infty}^{\infty} d^4l
=\frac{s}{2}\int\limits_{-\infty}^{\infty}d\alpha
\int\limits_{-\infty}^{\infty} d\beta \int\limits_{0}^{\infty}\pi dl^2_{\bot}.
\end{eqnarray}
The integration over the transverse momenta of the soft quark is performed
by taking half of the residues in the corresponding propagator 
\begin{eqnarray}
\int \frac{d^4l}{l^2-m^2+i0} F=\int\frac{\pi s}{2}
\frac{d\alpha d\beta d l^2_\bot}{s\alpha\beta-l_\bot^2-m^2+i0} F \to
-i\pi^2\frac{s}{2}\int d\alpha d\beta \Theta (s\alpha\beta-m^2)F.
\end{eqnarray}
In this manner the one loop amplitude can be calculated 
in DL approximation and result is, 
\begin{eqnarray}
F^{1-loop}=C\int\limits_{0}^{1}\int\limits_{0}^{1} 
\frac{d\alpha d\beta }{\alpha \beta} \Theta(\alpha \beta -\frac{m^2}{m_H^2} )
=C\Big( \frac12\ln^2 \frac{m^2}{m_H^2}\Big),
\end{eqnarray}
with $C=-\frac{4m^2}{m_H^2}.$   
 As mentioned earlier, the infrared sensitive contributions 
come from a region when the fermion line opposite 
the Higgs-fermion-fermion vertex is soft. 
This is, of course, also the origin of the term proportional to 
quark mass in the expression for the Higgs-$\gamma\gamma$ form
factor. In general, this one loop contribution gets radiative corrections and
the additional DL contributions arise from the region of soft gluons. It is well
known that to this accuracy they factorize and are independent of spin. Thus,
aiming at DL accuracy we can use the eikonal approximation, and easily generalize the above 
to include all orders in the QCD coupling. Let us briefly consider 
the resummation of these double logs for the Higgs-$\gamma\gamma$ 
form factor \cite{KoYa}. Some of the diagrams 
contributing to the next  order in $\alpha_s$ are shown in Figs.(1). 
Only the double logarithms arising from Fig.(1a) actually 
exponentiate -- other diagrams like those
in Figs.(1b, 1c) do not contribute as discussed below. 
With Fig.(1a) we note that we may
view the inner form factor as an off-shell Sudakov form 
factor (see Fig.2) with the "external" quark legs 
labelled $p_1, p_2$ (see Fig.1).  
\begin{eqnarray}\label{kinem2}
p_1^2=(k_1+k)^2=s\beta,\qquad p_2^2=(k-k_2)^2=-s\alpha.
\end{eqnarray}
Here, we are using the Sudakov parametrization for the momenta 
$k=\alpha k_1 +\beta k_2 + k_{\bot}$.

In order to implement the resummation (see Fig.2) we use the expression 
for the off-shell quark-anti-quark Sudakov 
form factor \cite{Poggio} 
\begin{eqnarray}\label{pg}
S(p_1 ,p_2)=\mbox{Exp}\Big( -\frac{C_F\alpha_s(\nu^2)}{2\pi}
\ln(\frac{ |p_1|^2}{s})\ln(\frac{ |p_2|^2}{s}) \Big).
\end{eqnarray}
The kinematical region of interest is restricted by the kinematics of
the one loop integral, and can be read off from eq. (\ref{kinem1}) and
eq.(\ref{kinem2}): 
\begin{eqnarray}\label{lim1}
m^2 \ll |p_1|^2, |p_2|^2 \ll s.
\end{eqnarray}
We have mentioned that diagrams like Figs.(1b, 1c) do not contribute to the
accuracy we are interested in. Indeed all such diagrams are not included in 
eq.(\ref{pg}). It is important to note that, diagrams, like Figs.(1b, 1c) 
and their counterparts in higher orders that cannot be included in the off-shell 
Sudakov form factor are irrelevant not
just to leading but also to the next to leading accuracy. To see this consider 
Fig.(1b): The soft fermion line is labelled with momenta $p_1+k-k_1$.  
Using this
fact together with $k_1^2=0$ and that for the infrared sensitive contribution, the
lines should be nearly on-shell, we observe that there are no large scales
$\sim m_H$ associated with the vertex correction in the Feynman gauge. Similarly
self-energy corrections cannot produce any large logarithms of the type
$\ln \frac{m^2}{m^2_H}$ in the Feynman gauge.

Now using Eq.(\ref{pg}) together with the above discussion, 
we have first to DL accuracy, the following for the resummation
 of the diagrams in Fig.(2),
\begin{eqnarray}
F&=&C
\int\limits_{0}^{1}\int\limits_{0}^{1} 
\frac{d\alpha d\beta }{\alpha \beta} 
\Theta(\alpha \beta -\frac{m^2}{s} )
\mbox{Exp}\Big( -\frac{C_F\alpha_s}{2\pi}
\ln |\alpha| \ln |\beta|  \Big).
\end{eqnarray}

We transform the  exponent into the power series 
and find that the integral of the $n-$th term will be of the form
\begin{equation}\label{integral}
\int\limits_0^1 d \xi_1 \int\limits^{1-\xi_1}_0 d \xi_2
\xi_1^{n+a}\xi_2^{n+b}=\frac{\Gamma (n+a+1)\Gamma (n+b+1)}{\Gamma (3+2n+a+b)}.
\end{equation}
The final result at DL accuracy reads
\begin{eqnarray}\label{dlresult}
F_{DL}&=&F^{1-loop}\sum\limits_{n=0}^{\infty}
\frac{2\Gamma (n+1)}{\Gamma (2n+3)}
\Big( -\rho \Big)^n,
\end{eqnarray}
with $\rho =\frac{C_F\alpha_s(\mu^2)}{2\pi}L^2, L=\ln\Big(\frac{m^2}{s}\Big)$. 
The index $n$ shows the order of the amplitude in $\alpha_s^n$. We can 
clearly identify the separate  contributions of the fixed orders in
$\alpha_s$. On the other hand, if $\rho $ is large all terms in the
series are important, giving altogether some analytic 
function $F_{DL}(\rho)$. This function is identified with 
a hyper-geometric function $_2F_2(1,1;2,\frac{3}{2};z)$, namely    
\begin{eqnarray}\label{dlr}
F_{DL}&=&F^{1-loop}\sum\limits_{n=0}^{\infty}
\frac{2\Gamma (n+1)}{\Gamma (2n+3)}
\Big( -\rho \Big)^n={}_2F_{2}(1,1;2,\frac{3}{2},-\frac{\rho}{4})
F^{{1-loop}},
\end{eqnarray}
we recall here that, in general, the function  ${}_2F_{2}(a,b;c,d;z)$ is
defined by a series  
\begin{eqnarray}
{}_2F_{2}(a,b;c,d;z)=\sum\limits^{\infty}_{k=0}\frac{(a)_k(b)_k}{(c)_k(d)_k}\frac{z^k}{k!}
\end{eqnarray}
Taking into account identities $(1)_k=k!$, 
$(\frac{3}{2})_k=2^{-2k}\frac{(2k+1)!}{k!}$, and $(2)_k=
\frac{\Gamma(2+k)}{\Gamma(2)}$ \cite{Prud}, we have  
\begin{eqnarray}
{}_2F_{2}(1,1;2,\frac{3}{2},-\frac{\rho}{4})=
\sum\limits^{\infty}_{k=0}\frac{(k!)^2(-\rho)^k}{(2k+1)!(k+1)!}=
\sum\limits_{k=0}^{\infty}\frac{2\Gamma (k+1)}{\Gamma (2k+3)}
\Big( -\rho \Big)^k.
\end{eqnarray}
This gives us the final DL result, eq.(\ref{dlr}). 
For large values of the parameter $\rho$ the function $F_{DL}$ 
has the following asymptotic,  
\begin{eqnarray}
F_{DL}(\rho)=\frac{2\ln (2\rho )}{\rho}F^{1-loop}.
\end{eqnarray}
We see that despite the fact that perturbation theory blows up at 
large $\rho$, the resummed result gives a smooth well defined function.

\section{Next-to-leading-logarithmic accuracy}
It is possible to develop this approach to achieve 
next-to-leading-logarithmic accuracy. 
For this we need an expression for the Sudakov form factor with NLL
accuracy. In fact, such  an analysis already exists in the literature 
\cite{Smilga} for the case when the two external fermion lines are off-shell 
by the same amount, i.e., $p_1^2=p_2^2=p^2$: 
\begin{eqnarray} \label{smilga}
S_{NLL}(p,p)=\mbox{Exp}\Big( -\frac{C_F\alpha_s(p^2)}{2\pi}
\ln^2(\frac{s}{|p|^2})+\frac{3C_F\alpha_s(p^2)}{4\pi}
\ln (\frac{s}{|p|^2}) \Big).
\end{eqnarray}
For our purposes, we need to
extend the analysis to take into account that $p_1^2 \neq p_2^2$. 
It is easy to see that for the region, eq.(\ref{lim1}), the proof of
factorization and exponentiation given in \cite{Smilga} goes through 
 with straightforward changes. The one major change involves the 
 normalization of the coupling. 
 We have studied one and two loop diagrams for the $Hq\bar q$
vertex with slightly off-shell quarks with the following result:
\begin{eqnarray}\label{slresult}
S_{NNL}(p_1,p_2)&=&Exp \Big(
-\frac{\alpha_s (\nu^2) C_F}{2\pi} \Big( \ln \frac{|p_1|^2}{s} \ln
\frac{|p_2|^2}{s}
 +\frac{3}{4}\ln \frac{|p_1|^2}{s}
+\frac{3}{4}\ln \frac{|p_1|^2}{s} \Big) \Big), 
\end{eqnarray}  
with the normalization of the coupling constant determined to be  
$\nu^2=\sqrt {|p_1^2| |p^2_2|}$. 
We show only the double and single IR logarithms in eq.(\ref{slresult}).  
In order to understand this normalization,   
we have to consider the diagram shown in Fig.(3), 
where we are keeping track of the $n_f$ dependent pieces only since they
are separately gauge invariant.      
Such diagrams can be accounted for by considering the following 
 gluon propagator
\begin{eqnarray}
D^{ab}_{\mu\nu}=-i\delta^{ab}(g_{\mu\nu}-\frac{k_\mu
k_\nu}{k^2})\frac{1}{k^2}
\frac{1}{1+\Pi (k^2)},
\end{eqnarray}
 where $\Pi (k^2)$ is the vacuum polarization by the gluon; 
at the one loop level it is simply $\Pi=\frac{\alpha_s\beta_0}{4\pi}\ln \Big
(\frac{k^2}{\mu^2}e^C\Big), \beta_0=11-\frac{2}{3}n_f$, $C$ being 
a scheme-dependent constant ($\overline{MS}$ scheme $C=-\frac{5}{3}$).  

The diagram Fig.3 corresponds to the first term in the expansion 
of the gluon propagator in $\alpha_s$. 
The $n_f$--part of this result is, as mentioned earlier, a 
gauge invariant part of the complete set of two loop diagrams. 

Because the effects of the running coupling gives only single
logarithmic terms it is enough to consider the remaining integrals 
to DL accuracy. 
Namely, we may trace  only the terms proportional to 
$\ln \frac{|p_1|^2}{s} \ln \frac{|p_2|^2}{s}$ from Fig.3.  
At DL accuracy the spin structure of the amplitude is simple, so
that one  needs to consider the scalar integral only,  
\begin{eqnarray}
I=1+\frac{\alpha_s(\mu^2) C_F}{2\pi}(2p_1p_2)
\int \frac{d^4k}{(2\pi)^4} \frac{1}{(p_1+k)^2}
\frac{1}{(p_2-k)^2}\Big(1-\frac{\alpha_s\beta_0}{4\pi}
\ln (\frac{k^2}{\mu^2})\Big) \frac{i}{k^2}.
\end{eqnarray}
To evaluate this, we consider a slightly more general integral
\begin{eqnarray}
J=i\int \frac{d^4k}{(2\pi)^4} \frac1{(p_1+k)^2}
\frac{1}{(p_2-k)}\frac{\mu^{2\Delta}}{(k^2)^{1+\Delta}},
\end{eqnarray} 
which after expanding in $\Delta$ will give us the desired integral in $I$. 
Using Feynman parameters, this  integral is reduced to 
\begin{eqnarray}
J=-\frac{1}{(4\pi)^2}\int\limits_{0}^{1}dy
\frac{1}{A\Delta}\mu^{2\Delta}\Big[ e^{-\Delta \ln (A+B)}- 
e^{-\Delta \ln (B)} \Big],
\end{eqnarray}
with
\begin{eqnarray}
A(y)=p_2^2y^2+2p_1p_2y+p_1^2,\\
B(y)=-2p_1p_2y-p_2^2y-p_1^2,\\
A(y)+B(y)=p_2^2y(-1+y).
\end{eqnarray}
The function $A(y)$ has two zeros, $y_\pm$: 
\begin{eqnarray}
A=p_2^2(y-y_+)(y-y_-),\quad
y_\pm=\frac{-2p_1p_2\pm\sqrt{(2p_1p_2)^2-4p_1^2p_2^2}
}{2p_2^2}.
\end{eqnarray}
For very small virtualities, $p_1^2, p_2^2 \to 0$, the roots are simplified to
$y_+=-\frac{p_1^2}{s}, y_-=-\frac{s}{p_2^2}$.
Expanding the integrand of $J$ in $\Delta$ up to second order we have 
\begin{eqnarray}
J&=&\frac{1}{(4\pi)^2}\int\limits_{0}^{1}dy\frac{1}{p_2^2(y-y_+)(y-y_-)}\mu^{2\Delta}
\Big( \ln\Big(\frac{p_2^2y(1-y)}{(2p_1p_2+p_1^2)y+p_1^2}\Big) \\
&+& \frac{\Delta}{2} \Big[ -\ln^2 (p_2^2y(1-y)) + \ln^2 ((2p_1p_2+p_2^2)y+p_1^2) \Big]
\Big).
\end{eqnarray}
The final integration over $y$ is simple, the result is
\begin{eqnarray}
J&=&\frac{1}{(4\pi)^2 2p_1p_2}
\Big( - \ln \frac{|p_1|^2}{s} \ln \frac{|p_2|^2}{s}+\frac{\Delta}{2} 
\ln \frac{|p_1|^2}{s} \ln \frac{|p_2|^2}{s} \ln
(\frac{|p_1^2||p_2^2|}{\mu^4})
\Big).
\end{eqnarray}
We see, that the first term in this equation reproduces the DL result 
from eq.(\ref{pg}) and eq.(\ref{slresult}).  It can be checked,
that the second term suggests the  normalization of the
coupling constant to be, $\nu^2=\sqrt{|p_1^2||p_2^2|}$.  
Indeed, returning to the integral $I$, we find
\begin{eqnarray}
I=1-\frac{\alpha_s(\mu^2) C_F}{2\pi}
\ln \frac{|p_1|^2}{s} \ln \frac{|p_2|^2}{s}
\Big[ 1-\frac{\alpha_s(\mu^2)\beta_0}{4\pi}
\ln\Big(\frac{\sqrt{|p_1^2||p_2^2|}}{\mu^2}\Big) 
\Big].
\end{eqnarray}
It is clear that the last logarithm, containing the $\beta_0$ term,  
can be absorbed into the running coupling, giving $\alpha_s(\nu^2)$ with  
the normalization point $ \nu^2=\sqrt{|p_1^2||p_2^2|} $. 
 The exponentiation of the integral $I$ will give us the final off-shell
 Sudakov exponent, eq.(\ref{slresult}). 


In order to get single logarithms in eq.(\ref{slresult}) we have to include
 the numerator and the spin structure. 
 We do not present the 
details of these calculations here. Instead we note, that all logarithms 
we have accounted for are of infrared origin, $s \gg p_1^2,p_2^2\to 0$.  
We do not show the UV logarithms which come as a result of the 
normalization of the quark mass, $m$,  
or the quark Yukawa coupling, $g_{Hq\bar q}$.  
These logarithms are of the form $\gamma \ln \frac {s}{\mu^2}$,
and are related to the anomalous dimensions  of the quark mass and the 
Yukawa coupling, $\gamma$, and can be traced separately from the IR
logs.  
Such terms can be omitted if the Yukawa coupling and related to it 
the quark mass in the leading order result are normalized at a large
scale $\mu^2=s$. 
The formula eq.(\ref{slresult}) reproduces the expression for the Sudakov 
form factor at non-equal virtualities at DL accuracy derived by  
Carrazone et. al.  in \cite{Poggio}, eq.(\ref{pg}),
as well as at NLL with equal virtualities $p^2=p_1^2=p_2^2$ derived 
by Smilga in \cite{Smilga}, eq.(\ref{smilga}).

In addition, the normalization point 
$\nu^2=\sqrt{|p_1^2||p_2^2|}$ that we find 
reproduces that of the NLL results 
with equal virtualities $p^2=p_1^2=p_2^2$ derived by Smilga in 
\cite{Smilga}, eq.(\ref{smilga}).

In our opinion, this scale,  $\nu^2=\sqrt{|p_1^2||p_2^2|}$,  
has a very transparent origin. 
The vertex of the interaction of a soft gluon with an off-shell quark ($p_1^2$)
is described by the coupling $g(p_1^2)$. In the situation of  
gluon-exchange between two quarks with different virtualities, 
we have an effective  coupling  $g(p_1^2)g(p_2^2)$.  
 Using the running of the coupling $g^2(\mu^2)=4\pi\alpha_s(\mu^2)$, at one loop level, 
$\alpha(\mu^2)=\alpha_s(\nu^2)/(1+\frac{\alpha_s\beta_0}{4\pi}\ln(\frac{\mu^2}{\nu^2}))$,
we will find that the effective coupling   $g(p_1^2)g(p_2^2)$ is
reduced to $\alpha_s(\sqrt{|p_1^2 ||p_2^2| })$. 
This coincides with our previous results.

As a next step, we include this form factor inside the one loop  
triangle diagram 
and calculate the last one loop integration with the  form
factor which now accounts for all large logarithms with NLL accuracy. 
The final result for the next-to-leading-logarithmic form 
factor reads
\begin{eqnarray}
F_Q=F_{DL}+F_{NLL},
\end{eqnarray}
with $F_{DL}$ from eq.(\ref{dlresult}) and 
\begin{eqnarray}\label{final}
F_{NLL}&=&\frac{1}{L}
F^{1-loop}\sum_{n=1}^{\infty} \frac{\Gamma (n+1)}{\Gamma (2n+2)}
(-\rho )^n\Big(3-\frac{\rho\beta_0}{C_F}\frac{n}{2n+2}\Big( 
\frac{n+1}{2n+3}+\frac{\ln (s/\mu^2)}{L}\Big) \Big),   
\end{eqnarray} 
with $\beta_0=11-\frac{2n_f}3$, $n_f$ is a number of light flavors.
Because the typical virtuality is harder than $(2m_b)^2$, we take 
number of light flavors to be $n_f=5$. 
 Because the Yukawa coupling and 
the related  quark mass in the leading order result  
are normalized at a large scale $\mu^2=s$,  we have to use  
\begin{eqnarray}
F^{1-loop}=-\frac{ 2 m(m^2_H)\cdot m}{m_H^2} \ln^2 
\frac{m^2}{m_H^2}
\end{eqnarray}
in $F_{DL}$, where $m(m^2_H)$ is the $\overline{MS}$ mass and 
$m$ is the pole mass. 
Note, that the second mass in $F^{1-loop}$ is the pole one - it comes
from the t-channel soft quark propagator of the sub-diagram $q\bar
q\to\gamma\gamma$.  
This amplitude does not have any anomalous dimensions and, 
therefore,  $\mu$ independent.   
 The difference between the running mass and the pole mass should be   
accounted only at ${\cal O}(\alpha^n_sL^{2n-1})$. 
  
Some comments on the derivation of the eq. (\ref{final}) are in order. 
 First,  we have used eq.(\ref{integral}) in the derivation. 
 Second, the first term, the factor 3, comes 
from single logarithmic terms (two last terms in eq.(\ref{slresult})), 
which are not related to the running of $\alpha_s$, whereas the 
last two terms in eq.(\ref{final}) are related to
the running of the coupling constant: 
 \begin{eqnarray}\label{run1}  
\beta_0\ln(\sqrt{|p_1^2||p_2^2|}/\mu^2)=\beta_0 (
\frac12 \ln (|\alpha|)+\frac12 \ln (|\beta|)+ \ln\frac{s}{\mu^2}). 
\end{eqnarray}
It is interesting, that if we chose  the normalization scale in eq.
(\ref{slresult})  
to be $k^2_\bot$, we would get a slightly different result. 
 Such a normalization has been used in \cite{MeStHqq}.  
 Using the fact that $k^2_\bot=\alpha\beta s$,  
the normalization scale in the Sudakov form factor for this choice 
of the normalization becomes $\mu^2=\frac{|p_1^2||p_2^2|}{s}$. 
Going to $F(\rho)$, the only difference in the calculations 
will enter through the term 
$\beta_0\ln(\frac{|p_1^2||p_2^2|}{s\mu^2})=\beta_0 (
\ln (|\alpha|)+ \ln (|\beta|)+ \ln\frac{s}{\mu^2})$ (compare with 
eq.(\ref{run1})). It would give a factor two larger 
non-logarithmic contribution proportional to $\beta_0$ 
in the final result for $F_Q$, eq.(\ref{final}).
 It is worth mentioning, that this scale,
$\mu^2=\frac{|p_1^2||p_2^2|}{s}$, does not agree with results of Smilga
at $p^2_1=p^2_2=p^2$, $\mu^2=p^2$ nor with our scale 
$\mu^2=\sqrt{|p_1^2||p_2^2|}$. 
We stress, that all our results, especially eq.(\ref{final}) 
are valid only for very large L, $|L| \gg 1$.

We may expand the expression for $F_Q$ at the two loop level.   
Redefining all masses through $m(\mu^2)$ by using 
\begin{eqnarray}
m=m(\mu^2)\Big(1+\frac{\alpha_sC_F}{2\pi}\Big[\frac{3}{2}\ln\frac{\mu^2}{m^2}
+2\Big]\Big),
\end{eqnarray}
we have  
\begin{eqnarray}
F_Q=-\frac{2m^2(\mu)}{m_H^2}\ln^2 \frac{m^2}{m_H^2}
\Big(1+\frac{C_F\alpha_s}{2\pi}\Big[ -\frac{1}{12}\ln^2
\frac{m^2}{m_H^2}
+\ln \frac{m^2}{m_H^2} + 3\ln \frac{\mu^2}{m^2}
\Big]\Big).
\end{eqnarray} 
This expanded result at two loops is in agreement with that  
in \cite{Spira3}, and can be viewed as a powerful check 
of our new resummed result. 

\section{Numerical results and discussions}

Having at hand all analytical NLL results for the form factor $F_Q$ 
we turn now to the numerical analysis.  
In order to get some estimates, we have used $m_b=4.5$ 
 GeV and the coupling constant
 normalized at $\alpha_s(m_Z)=0.118$ \cite{PDG}.
First, in Fig.4, we show the ratio of two amplitudes,  
$R(m,m_H,\mu )=\frac{F_Q(H\to \gamma\gamma)}{F_{1-loop}}$, 
as a function of the Higgs  mass.  
 The result of the purely double logarithmic resummation is 
presented at different normalization scales: at the soft scales  
$\mu^2=m_b^2, 9 m_b^2$ by the dashed-dotted and the short-dashed curves,  
at the hard scale $\mu^2= s$ by the dashed line, and at the 
intermediate scale $\mu^2=s x^{0.4}, x=m_b^2/m_H^2$ by the dotted curve.  
 We see that the DL resummed result depends substantially on $\mu$.  
In order to improve the stability in $\mu$  we may include the 
$\beta_0$ term, which has to make the $\mu$ dependence 
of the results smoother. We have checked that this is indeed the case.  
 As an example, we show the DL results plus the $\beta_0$--part 
of the NLL correction, by the solid line in Fig.4, choosing $\mu^2=s$.    
 We note that, omitting the non-$\beta_0$ part  
is not so meaningful and the $\beta_0$ term has 
to be treated in the same way as the other  
 NLL radiative corrections, specially, since as we will see later, 
 its effects are smaller than other NLL corrections.   

In the Fig.5, we present   
 $R=\frac{F_Q (H\to \gamma\gamma)}{F_{1-loop}}$ as a function of the
 Higgs mass (in GeV) to  DL and NLL accuracy. 
In this figure the result of the DL--resummation is shown by the dashed
 line and the result of the resummation to NLL accuracy
is shown by the solid line.  We see that the correction is moderate and
 positive. 
The substantial contribution comes from the non $\beta_0$ part as 
 mentioned previously.  
It is easy to understand the size and the sign of the DL and SL
 effects. 
In fact, the typical value for $\alpha_s\ln \frac{m^2}{m_H^2}\approx 1$, but the
 numerical factor $\frac{1}{2\pi}$ makes the parameter $\rho$ to be $\rho \approx
 0.1$. That is the size of the DL corrections.  
The relative size of the SL corrections   
in comparison to the DL contributions is 
estimated as $\frac{1}{L}$, so that the absolute correction  
is of order 5\%. We assume the range $m_H=100-500$ GeV for the Higgs
 mass.  The sign of SL and DL corrections in the Sudakov form factor are
 different, being negative for DL and positive for the SL in the
 exponent.  That in turn implies the positive sign for NLL effects in eq.(\ref{final}).   

 The normalization point in the final result for 
the Sudakov form factor is $\nu^2=\sqrt{|p_1^2||p_2^2|}$, which 
corresponds to some  $\mu^2$, such that  $m_b^2\ll \mu^2\ll s$. 
We see from eq.(\ref{final}) that the $\beta_0$ terms are zero at   
\begin{eqnarray}
\mu^2=s\Big(\frac{m_b^2}{s}\Big)^{a}
\quad\mbox{with}\quad a=\frac{n+1}{3n+2}.
\end{eqnarray} 
 The function $a(n)$ changes with $n$ in the interval from $\frac{1}{3}$  
up to $\frac{2}{5}$, so that the typical value of $a$ is about $a=0.4$. 
That is why the choice of $\mu^2= s x^{0.4}$ in the DL results, reproduces the 
DL result plus  the $\beta_0$ NLL result very well. 
Finally, we stress 
that the NLL corrections are only moderate  
and the substantial part comes   
from effects which are not related to the running of the coupling constant. 

\section{Conclusion}
In this letter we have studied the logarithmic QCD corrections to 
the Higgs coupling to two photons. 
We have developed a method for the resummation of large QCD corrections
in the form of Sudakov-like logarithms of the type 
$\alpha_s^n\ln^{2n,2n-1}({m \over m_H})$
which can contribute to this process in certain models, such as  
the MSSM for large $tan \beta$, up to
next-to-leading-logarithmic (NLL) accuracy. 
Our main result is eq.(\ref{final}).   
The NLL correction to the form factor is moderate, 
of order 5\%, the substantial part of which comes from 
terms not related to running coupling effects. 
 
The new type of QCD corrections in turn imply that there are 
additional uncertainties in the QCD correction to $H\to\gamma\gamma$ in 
the MSSM with large $tan\beta$.  
 This QCD correction is important only for the MSSM - it does not show up in 
the Standard Model where the light quark contributions are suppressed.
 
 Some of the ideas presented here can also be applied to the process 
$\gamma\gamma \to b\bar b$. The details will be published elsewhere.  

\section*{Acknowledgements}
We thank G. Kane, G. Korchemsky, M. Melles, K. Melnikov, M. Spira,  
S. Rigolin, E. Yao for helpful discussions. 
This work has been supported in part by the US Department of Energy. 

\newpage
\begin{figure}
\centerline{\epsfig{file=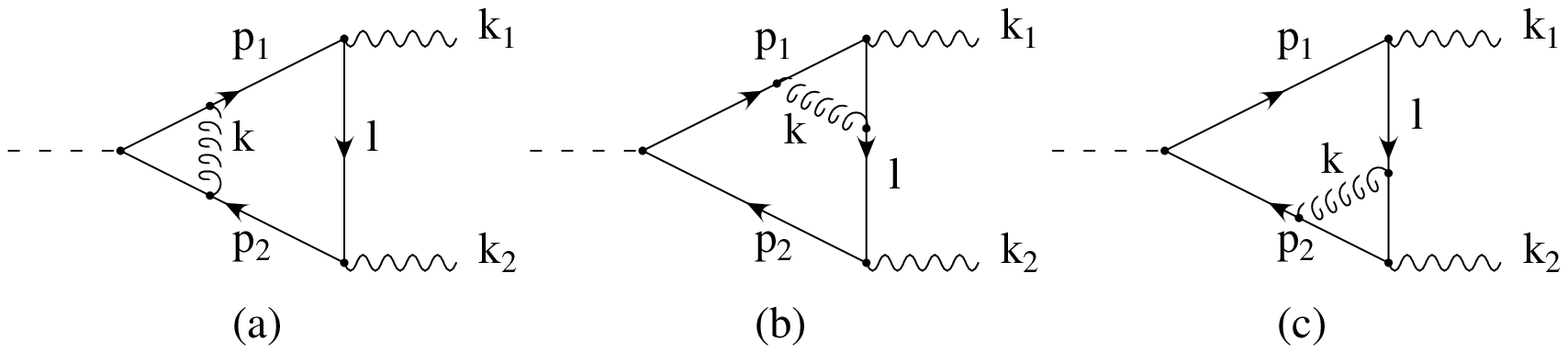,scale=0.85}}
\caption{\label{fig1} 
Some diagrams, which represented QCD corrections to the $H\to\gamma\gamma$
decay.}
\end{figure}
\begin{figure}
\centerline{\epsfig{file=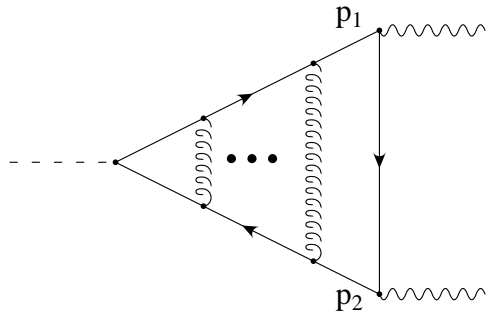,scale=0.85}}
\caption{\label{fig2} 
Diagrams contributing to DL and NLL in higher orders.}
\end{figure}
\begin{figure}
\centerline{\epsfig{file=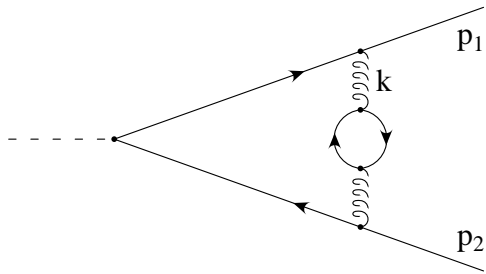,scale=0.85}}
\caption{\label{fig3} 
The diagram responsible for the normalization scale setting 
in the coupling constant.}
\end{figure}
\begin{figure}
\centerline{\epsfig{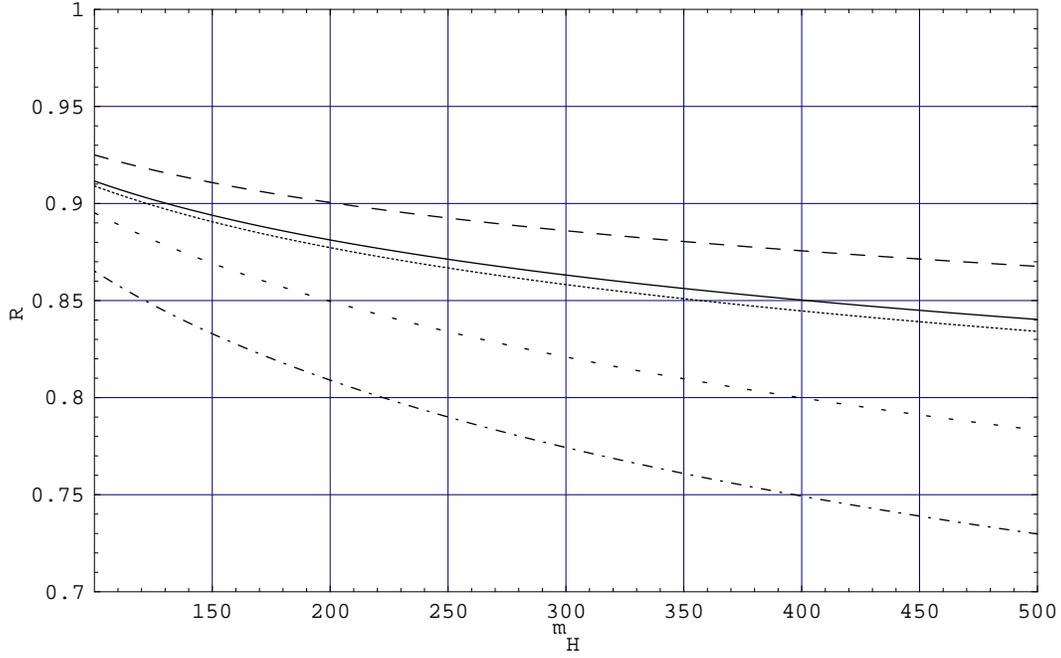}}
\caption{\label{fig4} 
The ratio $\frac{F_Q(H\to \gamma\gamma)}{F_{1-loop}}$ as a 
function of the Higgs mass. The result of DL resummation plotted 
at the different normalization scales $\mu$. 
The dotted-dashed curve corresponds to  $\mu^2=m^2_b$, 
the short-dashed curve to the $\mu^2=9m^2_b$, 
the dashed curve to $\mu^2=s$, the dotted curve  
to $\mu^2 =sx^{0.4}$. In addition, we show  
the DL result contribution  with the $\beta_0$ part of the NLL
correction at $\mu^2=s$ by the solid curve. }
\end{figure}
\begin{figure}
\centerline{\epsfig{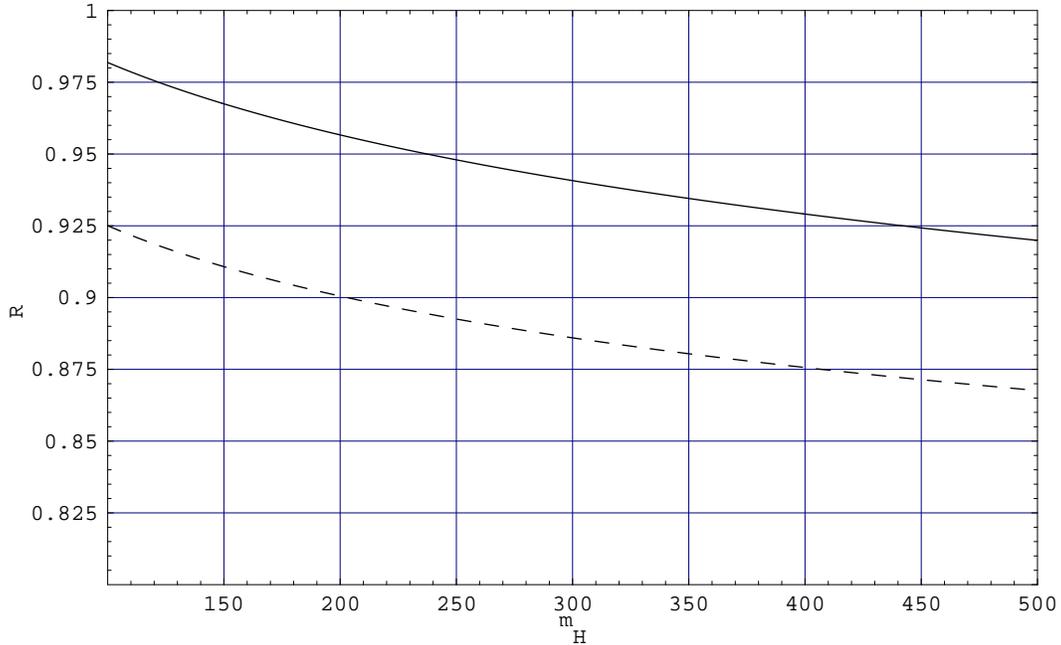}}
\caption{\label{fig5} 
The ratio $\frac{F_Q (H\to \gamma\gamma)}{F_{1-loop}}$ as a
function of the Higgs mass. The result of the DL resummation is shown 
by the dashed line and the the result of the resummation to NLL accuracy
is shown by the solid line, $\mu^2=s$.}
\end{figure}

\newpage
\begin{thebibliography}{9}
\bibitem{LEPHIGGS} 
A. Read, LEPC presentation on July 2000, 
http://lephiggs.web.cern.ch/LEPHIGGS.
\bibitem{Ginz1} I.F. Ginzburg et al, Nucl. Inst. Meth. 205 (1983) 47. 
\bibitem{Ginz2} I.F. Ginzburg et al, Nucl. Inst. Meth. 219 (1984) 5.
\bibitem{Telnov} V. I. Telnov, Nucl. Inst. Meth. A 355 (1995) 5.
\bibitem{MeYa1} K.~Melnikov and O.~Yakovlev,
Phys.\ Lett.\  {\bf B324}, 217 (1994).
\bibitem{MeYa2} J.~G.~Koerner, K.~Melnikov and O.~Yakovlev,
Phys.\ Rev.\  {\bf D53}, 3737 (1996).
\bibitem{ZeSp1}
A.~Djouadi, M.~Spira and P.~M.~Zerwas,
Phys.\ Lett.\  {\bf B311}, 255 (1993).
\bibitem{KoYa}
M.~I.~Kotsky and O.~I.~Yakovlev,
Phys.\ Lett.\  {\bf B418}, 335 (1998). 
\bibitem{MeStHqq}
M.~Melles,
Phys.\ Rev.\  {\bf D60}, 075009 (1999).
\bibitem{tanbeta}
M.~Carena {\it et al.},
``Report of the Tevatron Higgs working group,'' hep-ph/0010338.
\bibitem{MSSM} 
H.~E.~Haber and G.~L.~Kane, Phys. Rep. 11 (1985) 75.
\bibitem{HiggsRC} 
H.~E.~Haber,
``Higgs bosons in the minimal supersymmetric model: 
The Influence of radiative corrections,''
in ``Perspectives on Higgs Physics'' 
ed. G.L. Kane, World Scientific, Singapore (1993), p.79. 
\bibitem{FKM} V.~S.~Fadin, V.~A.~Khoze and A.~D.~Martin,
Phys.\ Rev.\  {\bf D56}, 484 (1997).
\bibitem{Ellis} J. Ellis, M. K. Gaillard, D. V. Nanopolous,
Nucl. Phys. B 106 292 (1976).
\bibitem{Shifman} 
M. Shifman et al,
Sov. J. Nucl. Phys. 30 (1979) 1368.
\bibitem{Sudakov} 
V.~V.~Sudakov,
Sov.\ Phys.\ JETP {\bf 3}, 65 (1956).
\bibitem{Gorshkov}
V.~G.~Gorshkov et al,
Sov.\ J.\ Nucl.\ Phys.\  {\bf 6}, 95 (1968).
\bibitem{Poggio} J. Carazzone, E. C. Poggio, H. R. Quinn, Phys. Lett. B
57, 161 (1975).
\bibitem{Prud} A. P. Prudnikov, Yu. A. Brychkov, O.I. Marichev  
``Integrals and Series'', Moscow , 1981.
\bibitem{Smilga} A. V. Smilga, Nucl. Phys. B161, 449 (1979).
\bibitem{PDG} 
C.~Caso {\it et al.},
Eur.\ Phys.\ J.\ {\bf C3}, 1 (1998).
\bibitem{Spira3}
M.~Spira, A.~Djouadi, D.~Graudenz and P.~M.~Zerwas,
Nucl.\ Phys.\ B {\bf 453}, 17 (1995).
\end{thebibliography}
\end{document}